\documentclass[twocolumn,showpacs,preprintnumbers,amsmath,amssymb]{revtex4}
\usepackage{graphicx}
\usepackage{dcolumn}
\usepackage{bm}

\begin{document}

\preprint{APS/123-QED}
\title{General stability criteria for inviscid rotating flow \footnote{This work is supported by the National Foundation of
Natural Science (No. 40705027), the Knowledge Innovation Program
of the Chinese Academy of Sciences (Nos. KZCX2-YW-QN514), and the
National Basic Research Program of China (No. 2007CB816004). }}
\author{Liang Sun }
\email{sunl@ustc.edu.cn; sunl@ustc.edu} \affiliation{
 School of Earth and Space Sciences,
 University of Science and Technology of China, Hefei 230026, China.\\
and LASG, Institute of Atmospheric Physics, Chinese Academy of
Sciences, Beijing 100029, China. }

\date{\today}
\begin{abstract}
  The general stability criteria of inviscid Taylor-Couette flows
with angular velocity $\Omega(r)$ are obtained analytically.
First, a necessary instability criterion for centrifugal flows is
derived as $\xi'(\Omega-\Omega_s)<0$ (or
$\xi'/(\Omega-\Omega_s)<0$) somewhere in the flow field, where
$\xi$ is the vorticitiy of profile and $\Omega_s$ is the angular
velocity at the inflection point $\xi'=0$. Second, a criterion for
stability is found as
$-(\mu_1+1/r_2)<f(r)=\frac{\xi'}{\Omega-\Omega_s}<0$, where
$\mu_1$ is the smallest eigenvalue. The new criteria are the
analogues of the criteria for parallel flows, which are special
cases of Arnol'd's nonlinear criteria. Specifically, Pedley's
cirterion is proved to be an special case of Rayleigh's criterion.
Moreover, the criteria for parallel flows can also be derived from
those for the rotating flows.  These results extend the previous
theorems and would intrigue future research on the mechanism of
hydrodynamic instability.

\end{abstract}

 \pacs{47.32.-y, 47.20.-k, 97.10.Gz} \maketitle

The instability of the rotating flows is one of the most
attractive problems in many fields, such as fluid dynamics,
astrophysical hydrodynamics, oceanography, meteorology, etc. Among
them, the simplest one is the instability of pure rotation flow
between coaxial cylinders, i.e., Taylor-Couette flow, which has
intensively been explored \cite{Chandrasekhar1961,Drazin2004}.

Two kinds of instabilities in inviscid rotating flow have been
theoretically addressed in the literatures. One is centrifugal
instability, which was first investigated by Rayleigh
\cite{Rayleigh1880,Drazin2004}. He derived the circulation
criterion for the inviscid rotating flows that a necessary and
sufficient condition for stability to axisymmetric disturbances is
that the square of the circulation does not decrease anywhere.
This criterion is also be stated as the Rayleigh discriminant
$\Phi\geq 0$ (see
Eq.(\ref{Eq:stable_taylorflow_RayleighDiscriminant}) behind), and
is always be used in astrophysical hydrodynamics. It is also
generalized to non-axisymmetric flows \cite{Paul2005jfm}. The
other is known as instability due to two-dimensional disturbances
in rotating flows, which is similar to the shear instability in
parallel flow. We call this instability as the shear instability
in rotating flow hereafter. For this instability, Rayleigh also
obtained a criterion, i.e., inflection point theorem in inviscid
rotating flows, which is the analogue of the theorem in parallel
flows \cite{Rayleigh1880}. Following this way, Howard and Gupta
\cite{Howard1962b} found a stability criterion for two-dimensional
disturbance in inviscid rotating flow. However, the theoretical
results remains scarce, due to the complexity of rotating flow.

Comparing the instability in the rotating flow with shear
instability in parallel flows, the criteria for parallel flows are
much more abundant. Fj\o rtoft \cite{Fjortoft1950} and Sun
\cite{SunL2007ejp,SunL2008cpl} proved some more strict criteria.
And the stability of two-dimensional in a rotating frame was also
addressed by Pedley \cite{Pedley1969}, which seems to be more
complex than the stability problem in the pure rotation flows.

Motivated, then, by the theoretical criteria for parallel flows
\cite{Fjortoft1950,SunL2007ejp}, our study focuses on the
instability due to shear in inviscid rotating flows. The aim of
this letter is to obtain such criteria for the inviscid rotating
flows, and the relationship between previous criteria is also
discussed. Thus other instabilities may be understood via the
investigation here.

For this purpose, Howard-Gupta equation (hereafter H-G equation)
\cite{Howard1962b} is employed. To obtain H-G equation, Euler's
equations \cite{Chandrasekhar1961,Drazin2004,CriminaleBook2003}
for incompressible barotropic flow in cylindrical polar
coordinates $(r,\theta)$ are then given by

 \begin{equation}
 \frac{\partial u_r}{\partial t} + u_r\frac{\partial u_r}{\partial
 r}+\frac{u_\theta}{r}\frac{\partial u_r}{\partial
 \theta}-\frac{u_\theta^2}{r}=-\frac{1}{\rho} \frac{\partial p}{\partial
 r},
 \label{Eq:stable_taylorflow_Radial}
 \end{equation}
and
 \begin{equation}
 \frac{\partial u_\theta}{\partial t} + u_r\frac{\partial u_\theta}{\partial
 r}+\frac{u_\theta}{r}\frac{\partial u_\theta}{\partial
 \theta}+\frac{u_ru_\theta}{r}=-\frac{1}{\rho r} \frac{\partial p}{\partial
 \theta}.
 \label{Eq:stable_taylorflow_Angular}
 \end{equation}
Under the condition of incompressible barotropic flow, the
evolution equation for the vorticity can be obtained from
Eq.(\ref{Eq:stable_taylorflow_Radial}) and
Eq.(\ref{Eq:stable_taylorflow_Angular}),
 \begin{equation}
\frac{\partial \xi}{\partial t} + u_r\frac{\partial \xi}{\partial
 r}+\frac{u_\theta}{r}\frac{\partial \xi}{\partial
 \theta}=0,
 \label{Eq:stable_taylorflow_Vorticity}
 \end{equation}
where $\xi=\frac{1}{r}\frac{\partial }{\partial
r}(ru_\theta)-\frac{1}{r}\frac{\partial u_r}{\partial \theta}$ is
the vorticity of the background flow.
Eq.(\ref{Eq:stable_taylorflow_Vorticity}) can also be derived from
Fridman's vortex dynamics equation
\cite{Batchelor1967,Saffman1992}. And it admits a steady basic
solution,
 \begin{equation}
 u_r=0, u_\theta=V(r)=\Omega(r) r,
 \label{Eq:stable_taylorflow_Basicflow}
 \end{equation}
where $\Omega(r)$ is the mean angular velocity. And Rayleigh
discriminant is defined by
 \begin{equation}
 \Phi=\frac{1}{r^3}\frac{d}{dr}(\Omega r^2)^2.
 \label{Eq:stable_taylorflow_RayleighDiscriminant}
 \end{equation}

Then, consider the evolution of two-dimensional disturbances. The
disturbances $\psi'(r,\theta,t)$, which depend only on $r$,
$\theta$ and $t$, expand as series of waves,
 \begin{equation}
\psi'(r,\theta,t)=\phi(r)e^{i(n\theta-\omega t)},
 \label{Eq:stable_taylorflow_Disturbance}
 \end{equation}
where $\phi(r)$ is the amplitude of disturbance, $n$ is real
wavenumber and $\omega=\omega_r+i\omega_i$ is complex frequency.
Unlike the wavenumber in Rayleigh's equation for inviscid parallel
flows, the wavenumber $n$ here must be integer for the periodic
condition of $\theta$. The flow is unstable if and only if
$\omega_i>0$. In this way, the amplitude $\phi$ satisfies
 \begin{equation}
(n\Omega-\omega)[D^*D-\frac{n^2}{r^2}]\phi-\frac{n}{r}(D\xi)\phi=0,
 \label{Eq:stable_taylorflow_HowardGuptaEq}
 \end{equation}
where  $D=d/dr$, $D^*=d/dr+1/r$. This equation is known as H-G
equation and to be solved subject to homogeneous boundary
conditions
\begin{equation}
D\phi=0 \,\, at\,\, r=r_1,r_2.
\label{Eq:stable_taylorflow_HowardGuptaBc}
\end{equation}
%


By multiplying $\frac{r\phi^{*}}{\omega-\Omega n}$ to H-G equation
Eq.(\ref{Eq:stable_taylorflow_HowardGuptaEq}), where $\phi^{*}$ is
the complex conjugate of $\phi$, and integrating over the domain
$r_1\leq r \leq r_2$, we get the following equation
\begin{equation}
\displaystyle\int_{r_1}^{r_2}
r\{\phi^*(D^*D)\phi-[\frac{n^2}{r^2}+\frac{nD(\xi)}{r(n\Omega-\omega)}]\|\phi\|^2\}dr\,=0.
\label{Eq:stable_taylorflow_HowardGupta_Inta}
 \end{equation}
Then the integration gives
\begin{equation}
\displaystyle\int_{r_1}^{r_2}
r\{\|\phi'\|^2+[\frac{n^2}{r^2}+\frac{n(\Omega
n-\omega^*)\xi'}{r\|\Omega n-\omega
\|^2}]\|\phi\|^2\}dr\,=0,
\label{Eq:stable_taylorflow_HowardGupta_Intb}
 \end{equation}
where $\phi'=D\phi$, $\xi'=D(\xi)$ and $\omega^{*}$ is the complex
conjugate of $\omega$. Thus  the real part and image part are
\begin{equation}
\displaystyle\int_{r_1}^{r_2}
r\{\|\phi'\|^2+[\frac{n^2}{r^2}+\frac{(\Omega-c_r)\xi'}{r\|\Omega-c\|^2}]\|\phi\|^2\}dr=0,
\label{Eq:stable_taylorflow_HowardGupta_Int_Rea}
 \end{equation}
and
\begin{equation}
\displaystyle\int_{r_1}^{r_2} \frac{c_i \xi'}{\|\Omega-c\|^2}\|\phi\|^2dr\,=0,
\label{Eq:stable_taylorflow_HowardGupta_Int_Img}
 \end{equation}
where $c=\omega/n=c_r+ic_i$ is the complex angular phase speed.
Rayleigh used only
Eq.(\ref{Eq:stable_taylorflow_HowardGupta_Int_Img}) to prove his
theorem: The necessary condition for instability is that the
gradient of the basic vorticity $\xi'$ must change sign at least
once in the interval $r_1<r<r_2$. The point at $r=r_s$  is called
the inflection point with $\xi'_s=0$, at which the angular
velocity of $\Omega_s=\Omega(r_s)$. This theorem is the analogue
of Rayleigh's inflection point theorem for parallel flow
\cite{Rayleigh1880,Drazin2004}.

Similar to the proof of Fj\o rtoft theorem \cite{Fjortoft1950} in
the parallel flow, we can prove the following criterion.

Theorem 1: A necessary condition for instability is that
$\xi'(\Omega-\Omega_s)<0$ (or $\xi'/(\Omega-\Omega_s)<0$)
somewhere in the flow field.

The proof of Theorem 1 is trivial, and is omitted here. This
criterion is more restrictive than Rayleigh's. Moreover, some more
restrictive criteria may also be found, if we follow the way given
by Sun \cite{SunL2007ejp}. If the velocity profile $\Omega(r)$ is
stable ($c_i=0$), then the hypothesis $c_i\neq0$ should result in
contradictions in some cases. So that a more restrictive criterion
can be obtained.

To find the criterion, we need estimate the rate of
$\int_{r_1}^{r_2} r\|\phi'\|^2 dr$ to $\int_{r_1}^{r_2} \|\phi\|^2
dr$,
\begin{equation}
\int_{r_1}^{r_2}r\|\phi'\|^2 dr=\mu\int_{r_1}^{r_2}\|\phi\|^2 dr,
\label{Eq:stable_taylorflow_Poincare}
\end{equation}
where the eigenvalue $\mu$ is  positive definition for $\phi \neq
0$. According to boundary condition
Eq.(\ref{Eq:stable_taylorflow_HowardGuptaBc}), $\phi$ can expand
as Fourier series. So the smallest eigenvalue, namely $\mu_1$, can
be estimated as $\mu_1>r_1\pi^2/(r_2-r_1)^2$
\cite{Mumu1994,SunL2007ejp}.

Then there is a criterion for stability using relation
(\ref{Eq:stable_taylorflow_Poincare}), a new stability criterion
may be found: the flow is stable if
$-(\mu_1+1/r_2)<\frac{\xi'}{\Omega-\Omega_s}<0$ everywhere.

To get this criterion, we introduce an auxiliary function
$f(r)=\frac{\xi'}{\Omega-\Omega_s}$, where $f(r)$ is finite at
inflection point. We will prove the criterion by two steps. At
first, we prove proposition 1: if the velocity profile is subject
to $-(\mu_1+1/r_2)<f(r)<0$, then $c_r\neq \Omega_s$.

Proof: Since $-(\mu_1+1/r_2)<f(r)<0$, then
\begin{equation}
   -(\mu_1+1/r_2)<\frac{\xi'}{\Omega-\Omega_s}\leq\frac{\xi'(\Omega-\Omega_s)}{(\Omega-\Omega_s)^2+c_i^2},
\end{equation}
and if $c_r=\Omega_s$ and $1\leq n$, this yields to
\begin{equation}
\begin{array}{rl} \displaystyle\int_{r_1}^{r_2}
r\|\phi'\|^2+[\frac{n^2}{r}+\frac{\xi'(\Omega-\Omega_s)}{\|\Omega-c\|^2}]\|\phi\|^2\,
dr &\geq \\
\displaystyle\int_{r_1}^{r_2}
[(\mu_1+\frac{1}{r_2})+\frac{1}{r}+\frac{\xi'}{(\Omega-\Omega_s)}]
\|\phi\|^2dr &>0.

\end{array}
\end{equation}
This contradicts
Eq.(\ref{Eq:stable_taylorflow_HowardGupta_Int_Rea}). So
proposition 1 is proved.

Then, we prove proposition 2: if $-(\mu_1+1/r_2)<f(r)<0$ and
$c_r\neq \Omega_s$, there must be $c_i^2=0$.

Proof: If $c_i^2\neq0$, then multiplying
Eq.(\ref{Eq:stable_taylorflow_HowardGupta_Int_Img}) by
$(c_r-c_t)/c_i$, where the arbitrary real number $c_t$ does not
depend on $r$, and adding the result to
Eq.(\ref{Eq:stable_taylorflow_HowardGupta_Int_Rea}), it satisfies
\begin{equation}
\displaystyle\int_{r_1}^{r_2}
r\{\|\phi'\|^2+[\frac{n^2}{r^2}+\frac{\xi'(\Omega-c_t)}{r\|\Omega-c\|^2}]\|\phi\|^2\}\,
dr=0.
\label{Eq:stable_taylorflow_Sun_Int} \end{equation}
But the above Eq.(\ref{Eq:stable_taylorflow_Sun_Int}) can not be
hold for some special $c_t$. For example, let $c_t=2c_r-\Omega_s$,
then there is $(\Omega-\Omega_s)(\Omega-c_t)<\|\Omega-c\|^2$, and
 \begin{equation}
\frac{\xi'(\Omega-c_t)}{\|\Omega-c\|^2}=
f(r)\frac{(\Omega-\Omega_s)(\Omega-c_t)}{\|\Omega-c\|^2}>-(\mu_1+\frac{1}{r_2}).
\label{Eq:stable_taylorflow_Sun_Ust}
 \end{equation}
This yields
\begin{equation} \int_{r_1}^{r_2}
[r\|\phi'\|^2+(\frac{n^2}{r}+\frac{\xi'(\Omega-c_t)}{\|\Omega-c\|^2})\|\phi\|^2]
dr>0,
\end{equation}
which also contradicts Eq.(\ref{Eq:stable_taylorflow_Sun_Int}). So
the second proposition is also proved.

Using 'proposition 1: if $-(\mu_1+1/r_2)<f(r)<0$ then $c_r\neq
\Omega_s$' and 'proposition 2: if $-(\mu_1+1/r_2)<f(r)<0$ and
$c_r\neq \Omega_s$ then $c_i = 0$', we find a stability criterion.

Theorem 2: If the velocity profile satisfy $-(\mu_1+1/r_2)<f(r)<0$
everywhere in the flow, it is stable.

This criterion is the analogue of the theorem proved by Sun
\cite{SunL2007ejp}. Both theorem 1 and theorem 2 here are more
restrictive than Rayleigh's theorem for the inviscid rotating
flows. The theorems indicate the probability that a vorticity
profile with local maximum or minimum would be stable, if it
satisfies the stable criteria. Theorem 2 implies that the rotating
flow is stable, if the distribution of vorticity is relatively
smooth. As shown by Sun \cite{SunL2007ejp}, the instability of
inviscid parallel flows must have vortices concentrated enough.
This is also the shear instability in rotating flows. Since
several stable criteria for inviscid rotating flows have been
obtained, it is convenient to explore the relationship among them,
as discussed followed.

The criteria for rotating flow can be applied to parallel flows,
given narrow-gap approximation. First, Pedley's criterion is
covered by the centrifugal instability criteria. As mentioned
above, Pedley \cite{Pedley1969} considered the stability of
two-dimensional flows $U$ in a frame rotating with angular
velocity $\Omega$. A criterion is found that instability occurs
locally when $2\Omega(2\Omega-U')<0$, where $U'=dU/dr$ represents
radial shear of horizontal velocity. Pedley's criterion, which is
recovered by later researches \cite{Tritton1992,Cambon1997}, is in
essence the special case of Rayleigh's circulation criterion,
i.e., $\frac{d}{dr}(\Omega^2r^4)<0$ for instability. Here the
proof is briefly given. Considering the narrow-gap approximation
$r_2-r_1=d\ll r_1$ and the large radii approximation
$1/r_1\rightarrow 0$ in Rayleigh's circulation criterion, there is
$\Omega'r\approx -U'$. So $\Phi=d(\Omega^2r^4)/{dr}/r^3=2\Omega
(2\Omega-U')<0$, which is exactly Pedley's criterion. Thus
Pedley's criterion is covered by Rayleigh's circulation criterion.
Second, the stable criteria for parallel flows, such as Rayleigh's
theorem \cite{Rayleigh1880} and Fj\o rtfot's theorem
\cite{Fjortoft1950}, can be derived from those for rotating flows,
given the narrow-gap approximation $r_2-r_1=d\ll r_1$ and the
large radii approximation $1/r_1\rightarrow 0$. Following this
way, the results of the Taylor-Couette system can also be applied
to the plane Couette system \cite{Faisst2000}. The proof is
omitted here, as the approach is trivial too. As pointed out by
Sun \cite{SunL2007ejp}, all of the shear instability criteria for
two-dimensional flows are the special cases of Arnol'd's nonlinear
criteria \cite{Arnold1969}, which are much more complex yet not
widely used. In general, all the known stability criteria for
parallel flows (even in a rotating frame) can be derived from the
stability criteria for rotating flows.

In summary, the general stability criterion is obtained for
inviscid rotating flow. These results extend Rayleigh's inflection
point theorem for curved and rotating flows, and they are
analogues of the theorems proved by Fj\o rtoft and Sun for the
two-dimensional inviscid parallel flows. Then Pedley's cirterion
is proved to be an special case of Rayleigh's criterion. Moreover,
the theorems for the parallel flows can be derived from those for
the rotating flows, given narrow-gap and large radii
approximations. These criteria extend the previous results and
would intrigue future research on the mechanism of hydrodynamic
instability.

\end{document}